\documentclass[twocolumns]{aa} 

\usepackage{graphicx}
\usepackage{txfonts}
\usepackage[switch]{lineno}

\usepackage{xcolor}

\newcommand{\kms}{{\mathrm{km~s^{-1}}}}

\newcommand{\pccm}{{\mathrm{pc~cm^{-3}}}}

\newcommand{\Modot}{{\mathrm{M_{\odot}}}}

\newcommand{\gcm}{{\mathrm{g~cm^{-3}}}}
\newcommand{\mus}{{\mathrm{\mu s}}}

\begin{document} 


\title{PSR~J1618$-$3921: a recycled pulsar in an eccentric orbit}

\titlerunning{PSR~J1618$-$3921: a recycled pulsar in an eccentric orbit}
\authorrunning{F. Octau et al.}


\author{
F.~Octau\inst{1} \and
I.~Cognard\inst{1,2} \and
L.~Guillemot\inst{1,2} \and
T.~M.~Tauris\inst{3,4} \and
P.~C.~C.~Freire\inst{3} \and
G.~Desvignes\inst{3}\and 
G.~Theureau\inst{1,2,5}
}

\institute{
\inst{1}~Laboratoire de Physique et Chimie de l'Environnement et de l'Espace -- Universit\'e d'Orl\'eans / CNRS, F-45071 Orl\'eans Cedex 02, France\\
\email{franck.octau@cnrs-orleans.fr}\\
\inst{2}~Station de radioastronomie de Nan\c{c}ay, Observatoire de Paris, CNRS/INSU, F-18330 Nan\c{c}ay, France\\
\inst{3}~Max-Planck-Institut f\"ur Radioastronomie, Auf dem H\"ugel 69, D-53121 Bonn, Germany\\
\inst{4}~Argelander-Institut f\"ur Astronomie, Universit\"at Bonn, Auf dem H\"ugel 71, D-53121 Bonn, Germany\\
\inst{5}~Laboratoire Univers et Th\'{e}ories (LUTh), Observatoire de Paris, CNRS/INSU, Universit{\'e} Paris Diderot, 5 place Jules Janssen, F-92190 Meudon, France\\
}

\date{Received XXX; XXX}

\abstract
{PSR~J1618$-$3921 is an $11.99$-ms pulsar in a $22.7$-d orbit around a likely low-mass He white dwarf companion, discovered in a survey of the intermediate Galactic latitudes at 1400 MHz conducted with the Parkes radio telescope in the late 1990s. Although PSR~J1618$-$3921 was discovered more than 15 years ago, only limited information has been published about this pulsar which has a surprisingly large orbital eccentricity ($e \simeq 0.027$), considering its high spin frequency and likely small companion mass.}
{The focus of this work is a precise measurement of the spin, astrometric and orbital characteristics of PSR J1618$-$3921. This was done with timing observations made at the Nançay Radio Telescope, from 2009 to 2017.}
{We analyzed the timing data recorded at the Nan\c{c}ay Radio Telescope over several years to characterize the properties of PSR~J1618$-$3921. A rotation ephemeris for this pulsar was obtained by analyzing the times of arrival of the radio pulses at the telescope.}
{We confirm the unusual eccentricity of PSR~J1618$-$3921 and discuss several hypotheses regarding its formation, in the context of other discoveries of recycled pulsars in eccentric orbits.}
{}


\maketitle

\section{Introduction} 

PSR~J1618$-$3921 is a Galactic-disk recycled pulsar discovered in a 1.4-GHz survey of the intermediate Galactic latitudes with the Parkes radio telescope \citep{Edwards_1618}. It is in a binary system with a likely low-mass He white dwarf (WD) companion; the orbit has a period of $22.7$ days. In the ``standard'' pulsar recycling scenario, millisecond pulsars (MSPs, here defined as pulsars with spin periods less than 15 ms) are spun-up by mass and angular momentum transfer from a companion star in a low-mass X-ray binary \citep{Alpar_1982, Radhakrishnan_1982, Bhattacharya_1991, Tauris_2006}. This process leads to highly circularized orbits, i.e., with very low eccentricities, $e$ \citep{Phinney1994}. Yet, early pulsar timing observations of PSR~J1618$-$3921 revealed an anomalously large eccentricity of about $0.027$ \citep{Bailes_1618}. This pulsar has thus been classified as an eccentric millisecond pulsar (eMSP defined here as MSPs with $e>0.01$), an emerging subclass of MSPs comprising a handful of objects (see Table \ref{table_emsp} and references therein). 
Interestingly, PSR~J1618$-$3921 and the other MSPs in this growing pulsar class display very similar orbital periods and eccentricity values (see Table \ref{table_emsp}), this suggests a common, non-chaotic formation process \citep{FreireTauris, antoniadis, Jiang}. These formation processes and the orbital parameters for PSR~J1618$-$3921 are discussed in detail in Section \ref{discussion}.

\begin{table*}[hbtp!]
\label{table_emsp}
\caption{Rotational period, orbital period, companion mass, orbital eccentricity and height above the Galactic plane for the six known eMSPs. Unlike other pulsars in this table, PSR J1903+0327 is believed to have been formed by triple-star interaction and is therefore isolated in the list. The quoted masses are the median masses derived from the mass function assuming a pulsar mass of 1.35 $M_{\odot}$ and a random distribution of orbital inclinations, except for PSR J1903+0327 and PSR J1946+3417, for which the pulsar and companion masses were determined through the measurement of post-Keplerian parameters. The height above the Galactic plane, $z$, was obtained from the ATNF pulsar catalogue \citep[see \protect\url{http://www.atnf.csiro.au/research/pulsar/psrcat,}][]{Manchester_2005}. The $z$ values all assume the YMW16 model of \citet{ymw16}.}
\begin{center}
\begin{tabular}{c c c c c c c}
   \hline
   PSR & $P$ (ms) &  $P_b$ (days) & $m_c~(\Modot)$ & $e$ & $|z|$ (kpc) & References \\
   \hline
   J0955$-$6150 & $2.00$ & $24.6$ & $0.22$ & $0.11$ & $0.22$ & \cite{Camilo_2015} \\
   J1618$-$3921 & $11.99$ & $22.7$ & $0.20$ & $0.027$ & $0.75$ & \cite{Edwards_1618, Bailes_1618} \\
   J1946$+$3417 & $3.17$ & $27.0$ & $0.2556(19)$ & $0.14$ & $0.57$ & \cite{Barr_2013,Barr_2017} \\
   J1950$+$2414 & $4.30$ & $22.2$ & $0.30$ & $0.08$ & $0.15$ & \cite{knispel_2015} \\
   J2234$+$0611 & $3.58$ & $32.0$ & $0.23$ & $0.13$ & $0.60$ & \cite{Deneva_2013,Antoniadis_2016} \\
   \hline
   J1903$+$0327 & $2.15$ & $95.2$ & $1.029(8)$ & $0.44$ & $0.1$ & \cite{crl+08,Freireal} \\
   \hline
 \end{tabular}
 \end{center}
 \end{table*}

Although several of the above-listed publications mentioned PSR~J1618$-$3921, a complete description of the timing properties and orbital characteristics of this pulsar has been lacking since its discovery more than a decade ago. We here present the results of the long-term timing of PSR~J1618$-$3921 with the Nan\c{c}ay decimetric Radio Telescope (hereafter NRT), confirming the pulsar's unusual eccentricity. The timing analysis is presented in Section~\ref{timing}. In Section~\ref{discussion} we discuss the main results, and different models proposed to explain the formation of PSR~J1618$-$3921 and other eMSPs. 


\section{Observations and data analysis}
\label{timing}

\begin{figure}[hbtp!]
\centering
\includegraphics[width=9cm]{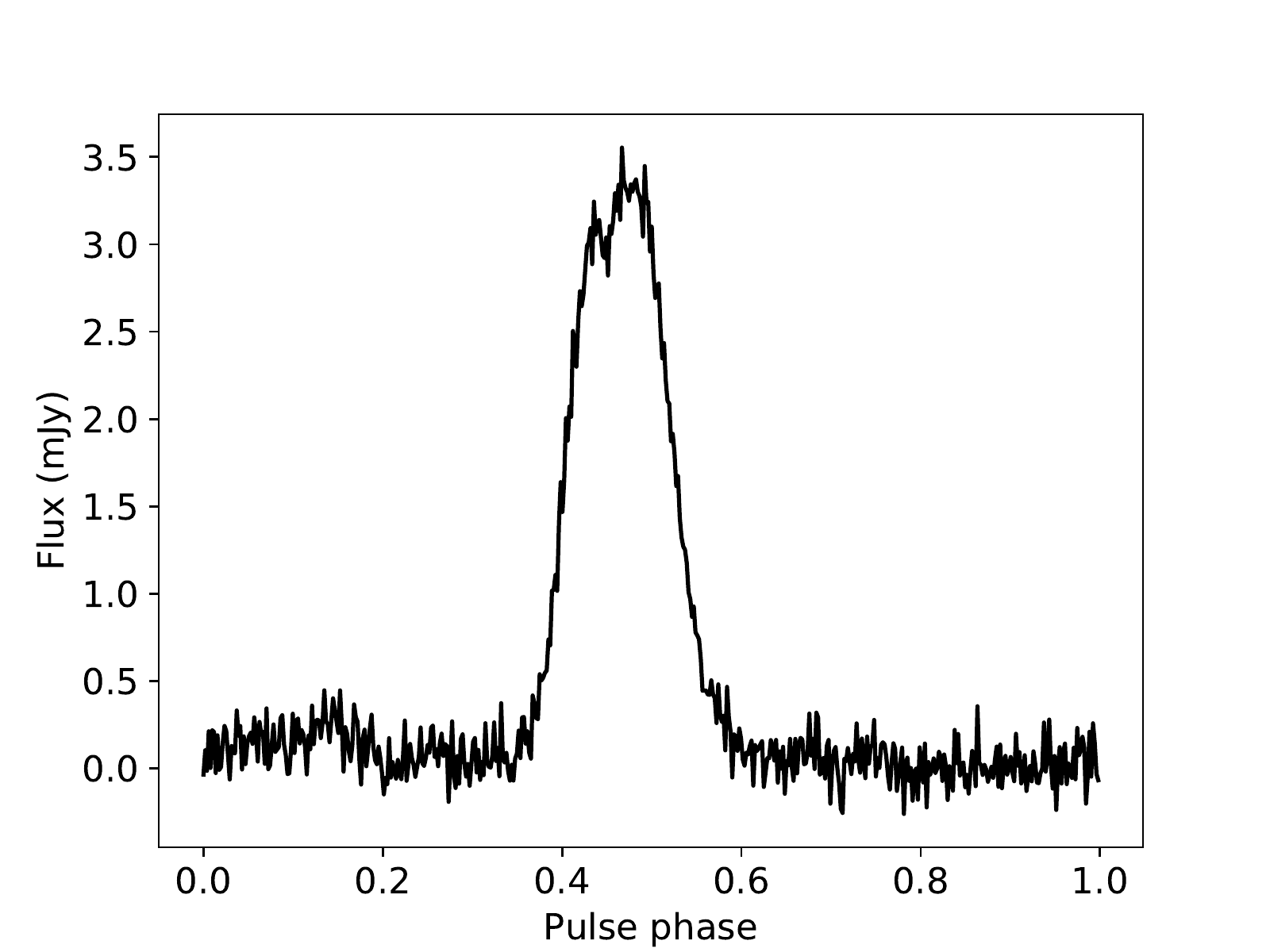}
\caption{Pulse profile for PSR~J1618$-$3921 at 1.4 GHz obtained by integrating 40 h of NRT observations conducted with the NUPPI backend in timing mode, between MJDs 56998 and 57869. The initial 2048 phase bins were divided by four in this graph to improve readability.}
\label{template}
\end{figure}

Observations of PSR~J1618$-$3921 with the NRT at 1.4 GHz began in May 2009. First observations were conducted with the Berkeley -- Orl\'{e}ans -- Nan\c{c}ay (BON) instrumentation, with very limited information regarding the pulsar's spin parameters, orbital parameters, and dispersion measure (DM). We therefore carried out observations with the BON backend in a so-called ``survey'' mode, i.e., with a 512 $\times$ 0.25 MHz incoherent filter bank sampled every 32 $\mu$s, giving us enough flexibility to search for the pulsar parameters in each observation and find its pulsations. After August 2011, pulsar observations at the NRT used the NUPPI instrumentation, a version of the Green Bank Ultimate Pulsar Processing Instrument\footnote{\url{https://safe.nrao.edu/wiki/bin/view/CICADA/NGNPP}} (GUPPI) designed for Nan\c{c}ay. At first, observations of PSR~J1618$-$3921 with NUPPI were also conducted in survey mode, with a total bandwidth of 512 MHz divided in 1024 channels of 0.5 MHz sampled every 64 $\mu$s. Survey mode NUPPI observations were conducted until a robust timing solution for PSR~J1618$-$3921 became available and enabled us to observe the pulsar in the regular ``timing'' mode, which we did starting from December 2014. In the regular timing mode of NUPPI, 128 channels of 4~MHz are coherently dedispersed in real time and the time series are folded on-line. In total, 10 survey mode observations of PSR~J1618$-$3921 were made with the BON instrumentation, and 60 were conducted with the NUPPI pulsar backend, including the first eight made in survey mode. Table~\ref{summary_obs} summarizes our observations of the pulsar, and Figure~\ref{template} displays a pulse profile for PSR~J1618$-$3921, obtained by summing all timing mode NUPPI observations. The profile comprises one main broad peak showing substructures, and evidence for a secondary peak leading the main one by approximately 0.35 rotation.

\begin{table*}[hbtp!]
\caption{Summary of observations of PSR~J1618$-$3921 with the NRT.}
\begin{tabular}{lccc}
\hline
Instrumentation & BON & NUPPI & NUPPI \\
\hline
Observation Mode & Survey & Survey & Timing \\
MJD range & $54963 - 55647$ & $56573 - 56946$ & $56998 - 57869$ \\
Calendar dates & May 2009 -- March 2011 & October 2013 -- October 2014 & December 2014 -- April 2017 \\
Number of observations & $10$ & $8$ & $52$ \\
Total observation time & 6h56min & 3h26min & 41h16min \\
Center frequency (MHz) & 1398 & 1484 & 1484 \\
Bandwidth (MHz) & 128 & 512 & 512 \\
Mean TOA uncertainty & $220~\mus$ & $193~\mus$ & $26~\mus$ \\
EFAC & $1.10$ & $0.67$ & $1.17$ \\
\hline
\end{tabular}
\label{summary_obs}
\end{table*}

The data reduction was done using the PSRCHIVE software library \citep{Hotan}. The data were cleaned of radio frequency interference (RFI) using a median smoothed automatic channel zapping algorithm. The polarization information was calibrated with the \textsc{SingleAxis} method of PSRCHIVE. For each observation configuration (see Table~\ref{summary_obs}) we combined all available 1.4 GHz observations and smoothed the resulting integrated profiles to form template profiles. The summed timing mode NUPPI profile at 1.4 GHz before smoothing is shown in Figure~\ref{template}. Finally, times of arrival (TOAs) were created by cross-correlating time- and frequency-scrunched observations with the template profiles. We obtained a dataset of 70 TOAs. The main properties of these TOAs are given in Table~\ref{summary_obs}.

\begin{figure*}[hbtp]
\centering
\includegraphics[scale=0.32]{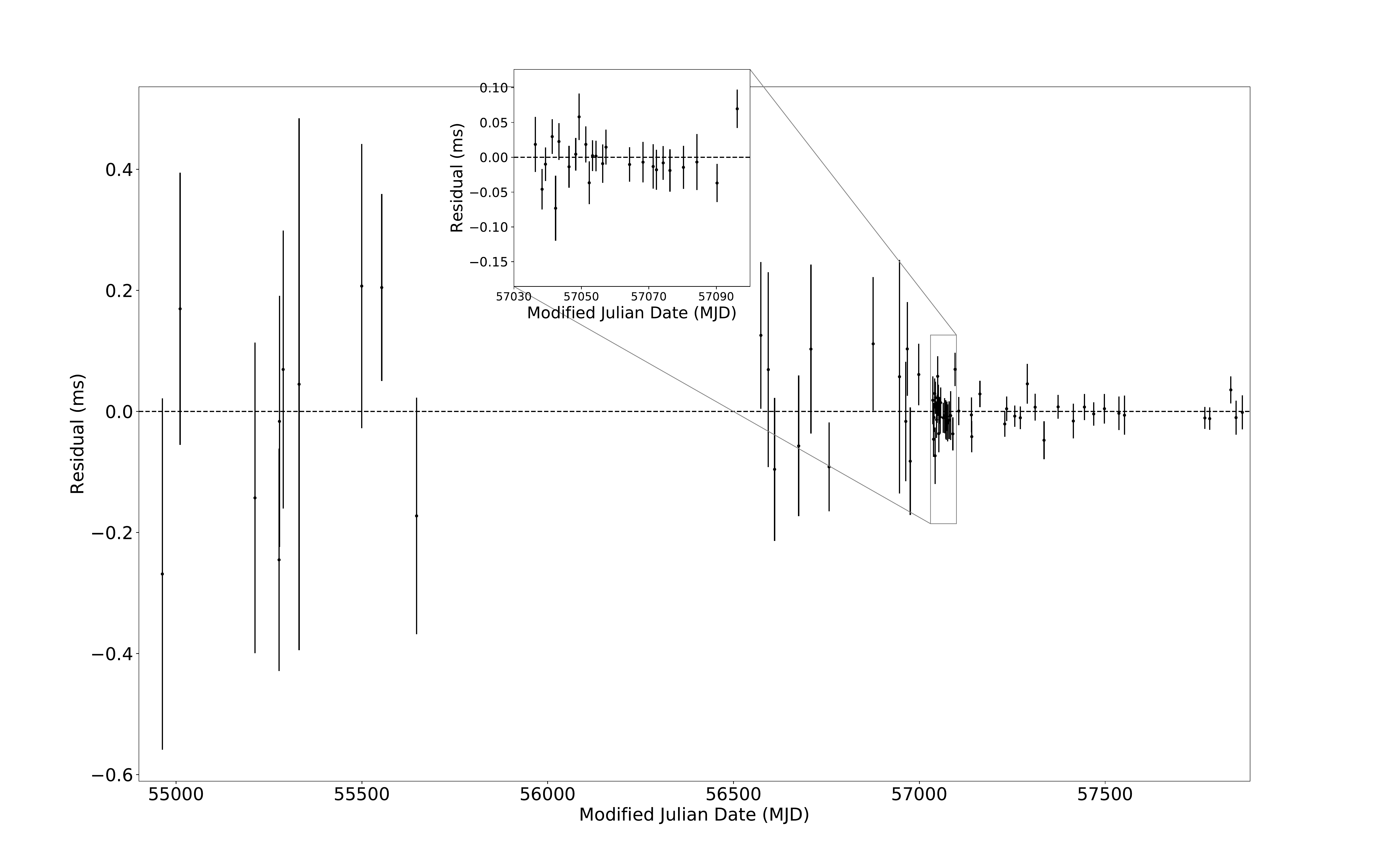}
\caption{Timing residuals as a function of time, for the best-fit timing solution of PSR~J1618$-$3921 given in Table~\ref{eph}. The inset shows the residuals for the time interval when the pulsar was intensively observed to derive a phase-connected timing solution.}
\label{residu}
\end{figure*}

The analysis of the TOA data was conducted with \textsc{Tempo2} \citep{Hobbs}. Because of the low signal-to-noise ratio (SNR) of the pulse profiles, in particular for the survey mode observations, we used fully frequency-scrunched TOAs in our timing analysis. Therefore, we could not determine the DM jointly with the other timing parameters. Nevertheless, in a first step of our analysis, we split timing mode NUPPI observations into four sub-bands of 128 MHz, and analyzed the TOAs with \textsc{Tempo2}, fitting for the DM and all other rotation parameters. We measured $\mathrm{DM} = 117.965 \pm 0.011\ \pccm$. In a last analysis step, we fixed the DM at the latter value, and analyzed the fully time- and frequency-scrunched TOAs. We fitted systematic time offsets between the different TOA datasets (i.e., BON survey, NUPPI survey and NUPPI timing) to account for differences in template profiles and observing systems and configurations. In addition, we fitted for the pulsar's sky position, rotational period and first time derivative, and five Keplerian parameters describing the orbit. The TOA uncertainties were weighted using \textsc{Tempo2} EFAC factors \citep{Hobbs}, to account for potential under- or over-estimations of the errors. Different EFAC factors were used for each of the TOA datasets; their values are listed in Table~\ref{summary_obs}. The best-fit parameters from the timing analysis are listed in Table~\ref{eph}, and the differences between the measured TOAs and the predictions of our best-fit model, the so-called residuals, are plotted in Figure~\ref{residu}. The weighted residuals have an RMS of $\sim 25.5$ $\mu$s, or about $2 \times 10^{-3}$ of the pulsar's rotational period. The reduced $\chi^2$ of approximately 1.2 indicates an acceptable agreement between the best-fit model and the data.

With the measured projected semi-major axis of the orbit, $x$, and orbital period, $P_b$, we derived the mass function given by:\\

\begin{equation}
 f(m_p, m_c) = \frac{(m_c\ sin\ i)^3}{(m_p + m_c)^2} = \frac{4 \pi}{T_{\odot}} \frac{x^3}{P_b^2} \; {= 0.00225347(9) \; M_{\odot}}
\end{equation}

\noindent
where $m_p$ is the pulsar mass, $m_c$ is the companion mass, $T_{\odot} = G\ M_{\odot} / c^3 = 4.925490947\ \mu$s and $i$ is the orbital inclination angle. Assuming an edge-on orbit ($i\ =\ 90^{\circ}$) and a pulsar mass of $1.35\ \Modot$, we can determine a lower limit on the companion mass $m_c$ of $\ 0.17\ \Modot$. As noted in ~\cite{lorimerkramer}, the probability of observing a binary system with an inclination of less than $i_0$ for a random distribution of orbital inclinations is $1-\cos(i_0)$; therefore, a 90\% confidence upper limit on the companion mass can be derived by assuming an inclination angle $i$ of $26^{\circ}$. Doing so gives an upper limit of $\ 0.44\ \Modot$ for the companion mass of PSR~J1618$-$3921 with a median mass of $\ 0.20\ \Modot$. The minimum mass of CO-type WDs is 0.33 M$_\odot$ \citep[e.g.][]{tlk2012}. Such a mass is only possible if the inclination angle is smaller than 34$^\circ$, which has a probability of $0.17$ assuming a random orbital orientation. Furthermore, the median companion mass fits the ($M_{\rm WD},P_{b}$)--correlation for He~WDs \citep{ts99} quite well. The value of the mass function and the range of likely companion mass values, combined with the orbital period, therefore indicate that the companion star is most likely a He-type WD. The orbital eccentricity $e$ of 0.0274339(10) is in line with the value reported by \citet{Bailes_1618}, confirming PSR~J1618$-$3921's unusually high orbital eccentricity, in spite of its common orbital period and companion type. Considering that the median spin period of the pulsars with He-WD companions is about 4 ms \citep[a value derived from pulsars listed in the ATNF pulsar catalogue\footnote{see \protect\url{http://www.atnf.csiro.au/research/pulsar/psrcat}},][]{Manchester_2005}, the spin period of PSR J1618$-$3921 appears to be an outlier. However, \cite{tlk2012} have pointed out half a dozen similar relatively slow spinning MSPs with apparent He-WD companions.

\begin{figure*}[hbtp]
\centering
 \includegraphics[width=6in]{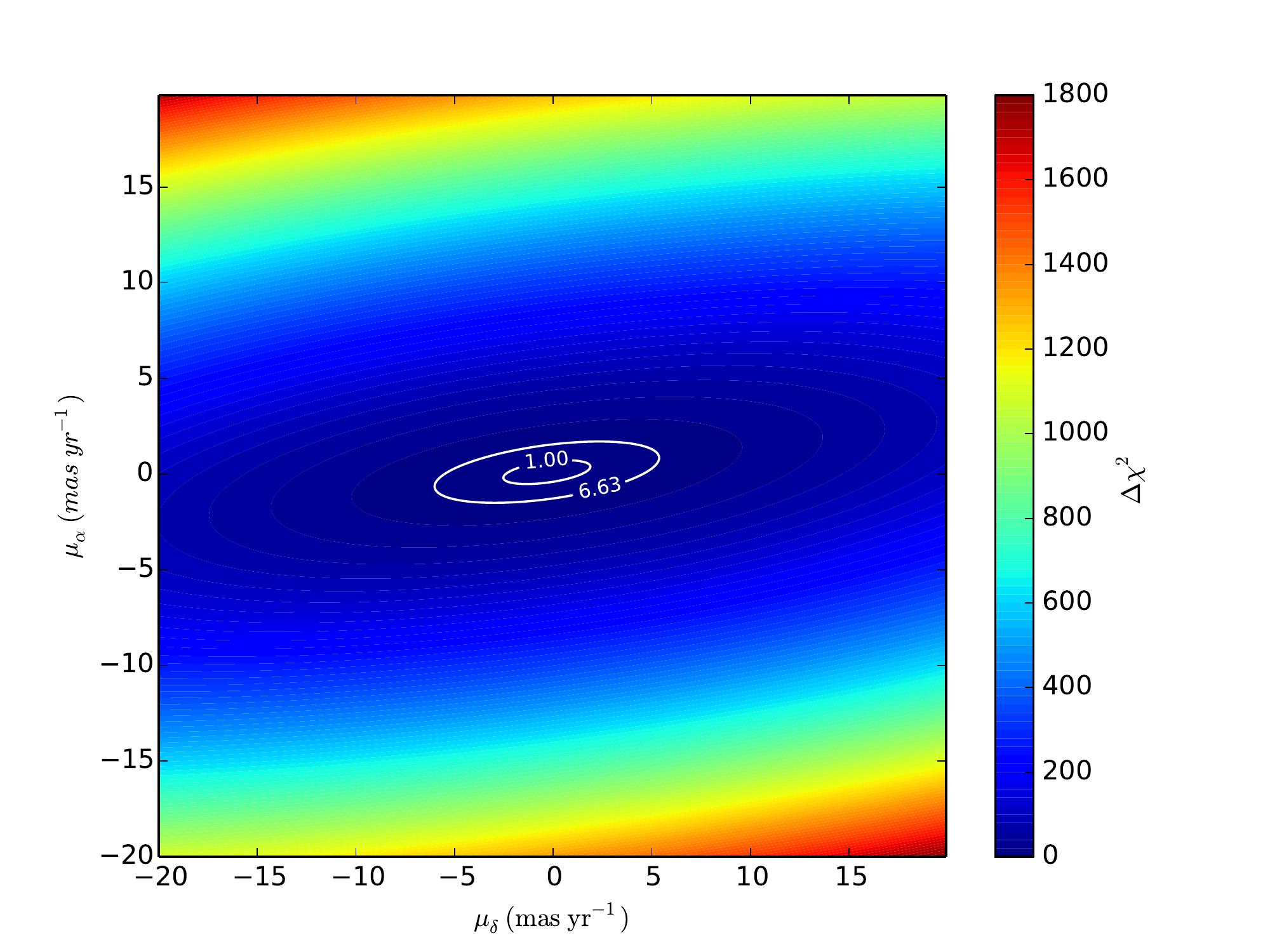}
\caption{Map of the $\chi^2$ variations as a function of the proper motions in right ascension and declination. Solid curves at $\Delta \chi^2 = 1.0$ and $6.63$ correspond to uncertainties of 1$\sigma$ and 3$\sigma$ respectively.}
\label{pmra_dec}
\end{figure*}

Because of the modest accuracy of the NRT timing for this low SNR pulsar, we were unable to measure any post-Keplerian parameters from which deriving further constraints on the masses of the binary. We also could not measure a significant timing parallax, which is unsurprising given the large DM distance estimates for this pulsar (see Table~\ref{eph}). However, we attempted to measure the pulsar's proper motion, since this is a key parameter for probing the formation process of eMSPs. As discussed in the following section, the different system formation scenarios predict varying initial kicks, so that the proper motion can be used to discriminate between them. A direct \textsc{Tempo2} fit did not lead to a significant proper motion measurement. We therefore produced a map of the $\chi^2$ of the timing residuals (see Figure~\ref{pmra_dec}), with proper motions in right ascension $\mu_\alpha$ and declination $\mu_\delta$ fixed at values ranging from $-20$ to $20$ mas yr$^{-1}$ with steps of 0.1 and 0.01 mas yr$^{-1}$ respectively, and fitting for the other parameters in Table~\ref{eph}. The minimum $\chi^2$ value is found for $(\mu_{\alpha}, \mu_\delta) = (0.10\ \mathrm{mas\ yr^{-1}}, -0.24\ \mathrm{mas\ yr^{-1}})$, but the 1$\sigma$ contour makes it consistent with 0. However, from the maximum transverse proper motion $\mu_T = \left(\mu_{\alpha}^2 + \mu_\delta^2\right)^{1/2}$ along the 3$\sigma$ contour (corresponding to $\Delta \chi^2 = 6.63$) we could determine an upper limit on PSR J1618$-$3921's transverse velocity of about 80 km s$^{-1}$, assuming the NE2001 model distance of 2.7 kpc. Assuming the YMW16 model distance of 5.5 kpc leads to a 3$\sigma$ upper limit on the velocity of 160 km s$^{-1}$. The two limits bracket the mean velocity of recycled pulsars of $\sim$90 km s$^{-1}$ \citep{hobbs_2005}, suggesting that PSR J1618$-$3921's transverse velocity could be low or typical, but not high.

\begin{table*}
\caption{Parameters for PSR~J1618$-$3921 derived from the timing analysis. The reference epoch for the position and for the period is MJD 56000. We used the DE421 Solar System ephemeris, time units are in barycentric dynamic time (TDB). Numbers in parentheses represent 1$\sigma$ \textsc{Tempo2} uncertainties in the last digits quoted. The DM value was estimated independently from the other parameters (see Section~\ref{timing}). The pulsar mass used to calculate the minimum companion mass is $1.35~M_{\odot}$. See \citet{lorimerkramer} for the calculation of $\tau$, $B$, $B_\mathrm{LC}$ and $\dot E$. Distance estimates were inferred from the NE2001 model of \citet{NE2001} and the YMW16 model of \citet{ymw16}. }
\centering
\begin{tabular}{ll}
\hline\hline
\multicolumn{2}{c}{Timing and binary parameters} \\
\hline
Right ascension, $\alpha$ (J2000)\dotfill &  16:18:18.8248(3) \\ 
Declination, $\delta$ (J2000)\dotfill & $-$39:21:01.815(10) \\ 
Rotational period, $P$ (ms) \dotfill & 11.987308585310(22) \\
Period derivative, $\dot{P}$ ($10^{-20}$) \dotfill & 5.408(18) \\
Dispersion measure, DM (cm$^{-3}$pc)\dotfill & 117.965(11) \\
Orbital period, $P_b$ (d)\dotfill & 22.74559403(19) \\ 
Projected semi-major axis of orbit, $x$ (lt-s)\dotfill & 10.278300(5) \\ 
Epoch of periastron, $T_0$ (MJD)\dotfill & 56012.21639(15) \\
Longitude of periastron, $\omega$ ($^{\circ}$)\dotfill & $-$6.717(3) \\ 
Orbital eccentricity, $e$\dotfill & 0.0274133(10) \\ 
Span of timing data (MJD) \dotfill & 54963.0---57869.1  \\  
Number of TOAs\dotfill & 70 \\
Weighted residual rms ($\mu s$)\dotfill &  25.3 \\ 
Reduced $\chi^2$ value \dotfill & 1.2 \\
\hline\hline
\multicolumn{2}{c}{Derived parameters} \\
\hline 
Mass function, $f$ ($M_{\odot}$) \dotfill & 0.00225347(9) \\
Minimum companion mass, $m_c$ ($M_{\odot}$) \dotfill & 0.1736 \\
Galactic longitude, $l$ ($^{\circ}$) \dotfill & 340.72 \\
Galactic latitude, $b$ ($^{\circ}$) \dotfill & 7.89 \\
Proper motion, $\mu_T$ (mas yr$^{-1}$) \dotfill & $<$ 6.0 \\
Characteristic age, $\tau$ ($10^{9}$ yr) \dotfill & 3.5 \\
Surface magnetic field strength, $B$ ($10^{8}$ G) \dotfill & 8.14 \\
Magnetic field strength at the light cylinder, $B_\mathrm{LC}$ ($10^{3}$ G) \dotfill & 4.4 \\
Spin-down luminosity, $\dot{E}$ ($10^{33}$ erg/s) \dotfill & 1.24 \\
Distance inferred from the NE2001 model, $d_\mathrm{NE2001}$ (kpc) \dotfill & 2.7 \\
Distance inferred from the YMW16 model, $d_\mathrm{YMW16}$ (kpc) \dotfill & 5.5 \\
\hline
\end{tabular}
\label{eph}
\end{table*}

\section{Discussion}
\label{discussion}

The large characteristic age ($3.5$ Gyr), relatively small surface magnetic field strength ($8.4\times 10^8$ G) and the small minimum companion mass ($0.17~M_\odot$) of PSR~J1618$-$3921 (see Table~\ref{eph}) are typical of MSPs formed by the recycling scenario \citep{tlk2012}. In such a scenario, the pulsar is spun-up by mass and angular momentum transfer from a companion star in a low-mass X-ray binary. Owing to a mass-transfer timescale much larger than the tidal-circularization timescale, the orbits of binary MSPs are highly circularized -- except for those systems formed in a globular cluster where the high stellar density may perturb the orbit. Considering that PSR~J1618$-$3921 does not belong to a globular cluster, the eccentricity of 0.027 is abnormal. We now discuss various hypotheses to explain such a high eccentricity.

\citet{FreireTauris} proposed the rotationally delayed, accretion-induced collapse (RD-AIC) of a super-Chandrasekhar mass WD. In this model, the binary system evolves from a system with a massive ONeMg~WD accreting from a low-mass companion star in a CV-like binary. After the mass transfer ceases, the system is composed of two WDs, a super-massive ONeMg~WD and a He~WD. As a result of its rapid spin, the ONeMg~WD was able to grow beyond the Chandrasekhar limit of $1.37\;M_{\odot}$ for a non-rotating WD \citep[][and references therein]{yl04}. The transition towards a neutron star--WD system occurs after the accretion phase when the ONeMg~WD loses spin angular momentum and the centrifugal forces are no longer sufficient to sustain the stability of the WD. It implodes and forms a neutron star. In this collapse process, the sudden release of gravitational binding energy increases the orbital period and imposes an eccentricity to the system. The observed properties of the resulting system do follow the ($M_{WD}$, $P_b$) relation \citep{ts99}. Monte Carlo simulations made by \citet{FreireTauris} reveal that to form a pulsar such as PSR~J1618$-$3921 (with an orbital period of 22.7 days and an eccentricity of 0.027), a small kick of $w\sim5\ \kms$ is necessary. Such small kicks are indeed expected from the similar electron capture supernovae \citep{kjh06}. In the simulations of \citet{FreireTauris} the masses of the ONeMg~WD and its He~WD companion were randomly chosen between $1.37 - 1.48\ \Modot$ and $0.26 - 0.28\ \Modot$, respectively. The RD-AIC model has the advantage to correctly account for some of the main observational and evolutionary characteristics of PSR~J1618$-$3921 (and of eMSPs in general). In particular: i) the observed mass of the He~WD companion fits the ($M_{\rm WD},P_{b}$)--correlation for He~WDs \citep{ts99}; ii) computations of the mass-transfer rate from its progenitor star (with its given orbital period) fall within the window where an ONeMg~WD can accrete and grow in mass beyond the Chandrasekar limit for non-rotating WDs \citep{tsyl13}; and iii) the predicted eccentricities fit the observed values. Nevertheless, the RD-AIC predicts small space
velocities and MSP masses which is in contradiction with the observation of at least one eMSP \citep[PSR~J1946+3417,][]{Barr_2017}. 

An alternative formation hypothesis was put forward by \citet{antoniadis} who demonstrated that the interaction of the binary with a circumbinary disk, fed by matter escaping the proto-WD during hydrogen shell flash episodes, may produce an eccentric system. For example, an interaction with a circumbinary disk with a typical mass of $ \approx (1-9) \times 10^{-4}\ \Modot$, and a lifetime of $10^4-10^5\;{\rm yr}$, is sufficient to produce an eccentric system with an eccentricity of $0.01 - 0.13$ and an orbital period of $15 - 50$ days. In this scenario, the eMSP would also follow the ($M_{WD}$, $P_b$) relation. Although the input physics remains uncertain \citep[e.g.][]{raf16}, the clear advantage of this formation scenario is that it does not impose any limitations on pulsar mass nor systemic velocity of the eMSP.

\citet{Jiang} proposed a model where a sudden mass loss, leading to the observed orbital eccentricity is caused by a phase transition in the pulsar interior due to quark deconfinement. This phase transition occurs when the core density of the neutron star reaches a critical density of $\sim 5 \rho_0$, where $\rho_0 = 2.7 \times 10^{14}\ \gcm$ is the nuclear saturation density. This process of quark deconfinement is not well known and the involved physics remains uncertain. 

Finally, a fourth possibility for producing eMSPs, is a triple star interaction in which one star is kicked out of the system, thereby leaving a fossil eccentricity to the remaining binary. This scenario is believed to have formed PSR~J1903+0327 \citep{crl+08,Freireal}. The unevolved, main-sequence G-type companion star in this system ($P_{b}=95\;{\rm days}$ and $e=0.44$) simply cannot have been responsible for the recycling of this MSP. However, as argued by \citet{FreireTauris} and \citet{knispel_2015}, for the remaining five eMSPs this triple-star process is too ``chaotic'' to produce quite identical observed properties for all systems, i.e. small eccentricities ($e\la 0.15$), similar orbital periods ($P_{b}\simeq 22-32\;{\rm days}$) and similar companion star masses ($M_2 \simeq 0.20-0.30\;M_{\odot}$). Indeed, simulations of triple-star interactions producing PSR~J1903+0327 confirm this argument \citep{pcp12}. Besides, the exchange scenario would not explain why eMSPs fit in the $(M_{WD}, P_b)$ law.

Better knowledge of the proper motion of all eMSPs is therefore an important tool to distinguish further between the formation scenarios for eMSPs. We need to collect more data in order to determine the proper motion of this pulsar more accurately. Early indications from our proper motion study seems to favour a small value and thus a low natal kick at the birth of the neutron star.

We note, in closing, the discovery by \citet{Keane_2018} of PSR~J1421$-$4409, an MSP with an orbital period of $\sim$30-d and a ``normal'' eccentricity of about $10^{-5}$. This binary pulsar demonstrates that, whatever mechanism causes the anomalously large eccentricity of eMSPs, it does not affect all binary systems in this orbital period range.


\section{Conclusion \& future work}

The analysis of eight years of NRT data enabled us to obtain a fully coherent timing solution for the first time, and to confirm the anomalous eccentricity of PSR J1618$-$3921 ($e = 0.027$) first announced by \citet{Bailes_1618}. This, the orbital period and mass function of the system make it a member of an emerging class of eccentric binary MSPs with similar orbital characteristics for which the formation process is poorly understood. We place an upper limit on the pulsar's proper motion. An actual measurement of the proper motion, which would provide us with crucial information about the system's formation, was not possible with the considered dataset. Continued timing observations with the NUPPI should enable us to measure the pulsar's proper motion in a few years. In parallel, optical and/or infrared observations of the binary system would be beneficial, as they would provide important information about the companion star, expected from the timing analysis to be a He WD. Further discoveries of similar systems exhibiting anomalously large eccentricities would also be useful for understanding the formation of these peculiar binaries.

\begin{acknowledgements}

The Nan\c{c}ay Radio Observatory is operated by the Paris Observatory, associated with the French Centre National de la Recherche Scientifique (CNRS). We acknowledge financial support from ``Programme National de Cosmologie et Galaxies'' (PNCG) of CNRS/INSU, France. PCCF gratefully acknowledges financial support by the European Research Council for the ERC Starting grant BEACON and continued support from the Max Planck Society under contract No. 279702.

\end{acknowledgements}

\bibliography{octau}
\bibliographystyle{aa}

\end{document}